\documentclass{appolb}
\usepackage{graphicx}
\usepackage{epsfig}

\newcommand{\Pomeron}{{I\!\!P}}
\newcommand{\Reggeon}{{I\!\!R}}
\begin{document}
\title{Inclusive and diffractive dijet photoproduction at the Electron-Ion Collider in NLO QCD%
\thanks{Presented at XXIX Cracow EPIPHANY Conference on Physics at the Electron-Ion Collider and Future Facilities, Cracow, Poland, 
January 16-19, 2023}%
}
\author{V. Guzey
\address{University of Jyvaskyla, Department of Physics, P.O. Box 35, FI-40014 University of Jyvaskyla, Finland and Helsinki Institute of Physics, P.O. Box 64, FI-00014 University of Helsinki, Finland}
\\[3mm]
{M. Klasen
\address{Institut f\"ur Theoretische Physik, Westf\"alische Wilhelms-Universit\"at M\"unster, Wilhelm-Klemm-Stra{\ss}e 9, 48149 M\"unster, Germany}
}
}
\maketitle

\begin{abstract}

In the framework of collinear factorization and next-to-leading order (NLO) perturbative QCD, we make predictions for 
inclusive and diffractive dijet photoproduction in electron-proton and electron-nucleus scattering in the EIC kinematics.
We establish kinematic ranges in the ${\bar p}_T$, ${\bar \eta}$, $x_A^{\rm obs}$ and $x_{\gamma}^{\rm obs}$
variables, quantify sensitivity to small-$x$ nuclear PDFs, and analyze various scenarios of factorization breaking in the case of diffractive scattering.

\end{abstract}

\vspace*{-115mm} {\hfill MS-TP-23-08} \vspace*{100mm}
  
\section{Introduction}
\label{sec:into}

All currently available information on jet photoproduction on hadrons comes from electron (positron)-proton scattering  
at Hadron-Electron Ring Accelerator (HERA), for reviews, see~\cite{Butterworth:2005aq,Klein:2008di,Newman:2013ada}.
Provided that the jet transverse momenta $p_T$ are sufficiently large, this process allows one to probe the microscopic quark-gluon structure 
of the proton and the real photon in quantum chromodynamics (QCD) as well as the strong interaction dynamics in the regime of perturbative QCD (pQCD).  The predictions of next-to-leading order (NLO) pQCD provide a good description of the dijet photoproduction cross section measured at HERA as a function of various jet observables in a wide range of $p_T$~\cite{H1:2002apm,H1:2006rre,ZEUS:2001zoq,ZEUS:2007njl}.
%~\cite{H1:2002apm,H1:2006rre,ZEUS:2001zoq,ZEUS:2007njl}.
This serves as an important test of the QCD factorization and universality of parton distribution functions (PDFs).

A related important incentive to study photoproduction of jets is that the cross section of this process 
has enhanced sensitivity to the gluon distribution.
As a result, QCD analyses of the combined data on the dijet cross section and
the total cross section of
lepton-proton deep inelastic scattering (DIS) 
provide additional constraints on the gluon PDF of the proton, see, e.g.~\cite{ZEUS:2005iex}.
Similarly, the combination with the available data on the $F_2^{\gamma}(x,Q^2)$ photon structure function measured in
electron-positron annihilation enables one to better constrain the gluon PDF of the real photon~\cite{Slominski:2005bw}.
Also, in the case of diffractive dijet photoproduction, one can use this process to analyze the poorly understood mechanism of the QCD factorization breaking in
diffractive scattering observed experimentally~\cite{H1:2010xdi,H1:2015okx,ZEUS:2007uvk}.
%~\cite{H1:2007jtx,H1:2010xdi,H1:2015okx,ZEUS:2007uvk}. 

It is expected that studies of photoproduction of jets will be continued at the future Electron-Ion Collider (EIC) in the U.S.~\cite{Accardi:2012qut}
and the Large Hadron Electron Collider (LHeC)~\cite{LHeCStudyGroup:2012zhm} and/or a Future Circular Collider (FCC)~\cite{FCC:2018byv} at CERN.
It will allow one not only to measure this process in a kinematic region complimentary to that covered by HERA and with much higher precision, but 
will also give for the first time an access to novel nuclear diffractive PDFs in the case of nuclear beams.

Note that first results on inclusive dijet photoproduction on heavy nuclei have recently been obtained by ATLAS~\cite{ATLAS:2017kwa} by analyzing lead-lead
ultraperipheral collisions (UPCs) at the Large Hadron Collider (LHC). It was shown in~\cite{Guzey:2018dlm} that NLO pQCD provides a good 
description of these data.

\section{Inclusive dijet photoproduction in $eA$ scattering at EIC}
\label{sec:inc}

\begin{figure}[htb]
\centerline{%
\includegraphics[width=11cm]{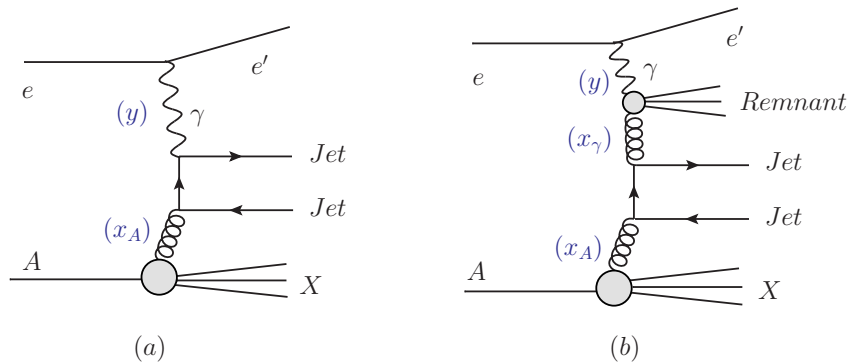}}
\caption{Typical LO direct-photon (left) and resolved-photon (right) contributions to dijet photoproduction in $eA$ scattering.
The involved momentum fractions $y$, $x_A$, and $x_{\gamma}$ are shown in parenthesis. }
\label{fig:photo_dijets_eA}
\end{figure}

As we explained in the Introduction, photoproduction of jets provides complementary information on the partonic structure of hadrons and the photon in QCD.
In particular, the process of inclusive dijet photoproduction in lepton-nucleus ($eA$) scattering, $e+A \to e^{\prime}+{\rm 2\,jets}+X$, is expected to yield new
constraints on nuclear PDFs. Typical leading order (LO) Feynman graphs for this process are shown in 
Fig.~\ref{fig:photo_dijets_eA}: graphs $(a)$ and $(b)$ represent the so-called direct-photon and the resolved-photon contributions, respectively. In graph $(a)$, the photon enters the hard process of the photon-gluon fusion directly as an elementary particle. In contrast, in graph $(b)$, the photon participates in hard scattering by means of its partonic content, which is hence revealed (resolved) 
in this process. 
%Note that while the separation between the direct and resolved photons is not unique beyond LO, it is still useful. 

In the framework of collinear factorization and NLO pQCD, the $e+A \to e^{\prime}+{\rm 2\,jets}+X$ cross section can be written as the following convolution~\cite{Frixione:1997ks,Klasen:2002xb}
\begin{eqnarray}
d\sigma(e+A &\to& e^{\prime}+ 2\,{\rm jets}+X)=\sum_{a,b}\int dy \int dx_{\gamma} \int dx_A f_{\gamma/e}(y) \nonumber\\
&\times& f_{a/\gamma}(x_{\gamma},\mu^2) f_{b/A}(x_A,\mu^2) d{\hat \sigma}(ab \to {\rm jets}) \,,
\label{eq:cs_inc}
\end{eqnarray}
where $f_{\gamma/e}(y)$ is the photon flux of the electron with $y$ being the momentum fraction carried by the photon;
$f_{a/\gamma}(x_{\gamma},\mu^2)$ are the photon PDFs in the resolved-photon case, which depend on the parton-in-photon
momentum fraction $x_{\gamma}$ and the scale $\mu$; $f_{b/A}(x_A,\mu^2)$ are nuclear PDFs depending on the parton
momentum fraction $x_A$ and the scale $\mu$; $d{\hat \sigma}(ab \to {\rm jets})$ is the cross section of hard scattering of partons 
$a$ and $b$ into jets.
In the direct-photon case, parton $a$ corresponds to be the photon leading to $f_{\gamma/\gamma}(x_{\gamma},\mu^2)=\delta(1-x_{\gamma})$ at LO.
In Eq.~(\ref{eq:cs_inc}), all involved hard scales 
%vg (the renormalization scale in the hard cross section, the factorization scales on the nucleus and the photon sides) 
%
have been set to be equal. In our analysis, we identify them with the mean transverse momentum of the two jets,
$\mu={\bar p}_T=(p_{T,1}+p_{T,2})/2$.
Note that while the separation between the direct and resolved photons is not unique beyond LO, it is still useful since
the direct-photon contribution peaks in the $x_{\gamma} \to 1$ limit.

Our predictions~\cite{Guzey:2020zza} for the dijet photoproduction in $eA$ scattering at the EIC are based on the numerical implementation of Eq.~(\ref{eq:cs_inc}) combined with the anti-$k_{T}$ jet clustering algorithm with at most 2 partons in a jet, which was developed in~\cite{Klasen:1995ab,Klasen:1996it,Klasen:1997br}. 
While the parton momentum fractions $x_{A}$ and $x_{\gamma}$ are not directly measurable, they can be approximated using
the following hadron-level estimates based on the jet transverse momenta $p_{T,1}$ and $p_{T,2}$ and the jet
(pseudo)rapidities $\eta_1$ and $\eta_2$,
\begin{eqnarray}
x_{A}^{\rm obs} &=& \frac{p_{T,1}e^{\eta_1}+p_{T,2}e^{\eta_2}}{2 E_A} \,, \nonumber\\
x_{\gamma}^{\rm obs} &=& \frac{p_{T,1}e^{-\eta_1}+p_{T,2}e^{-\eta_2}}{2 y E_e} \,,
\label{eq:hadron-level}
\end{eqnarray}
where $E_A$ and $E_e$ are the energies of the nucleus and electron beams, respectively.
For definiteness, we take $E_A=100$ GeV per nucleon and $E_e=21$ GeV corresponding to $\sqrt{s}=92$ GeV~\cite{Accardi:2012qut}. For final-state jets, we assume generic conditions based on 
the HERA experience: the leading jet has $p_{T,1} > 5$ GeV and the subleading jets carry $p_{T,i\neq 1} > 4.5$ GeV;
all jets have $\eta_{1,2} < 4$; the jet cone parameter is $R=0.4$.
Finally, we use the GRV HO photon PDFs~\cite{Gluck:1991jc} and the nCTEQ15 nuclear PDFs~\cite{Kovarik:2015cma}.

\begin{figure}[htb]
\centerline{%
\includegraphics[width=6.5cm]{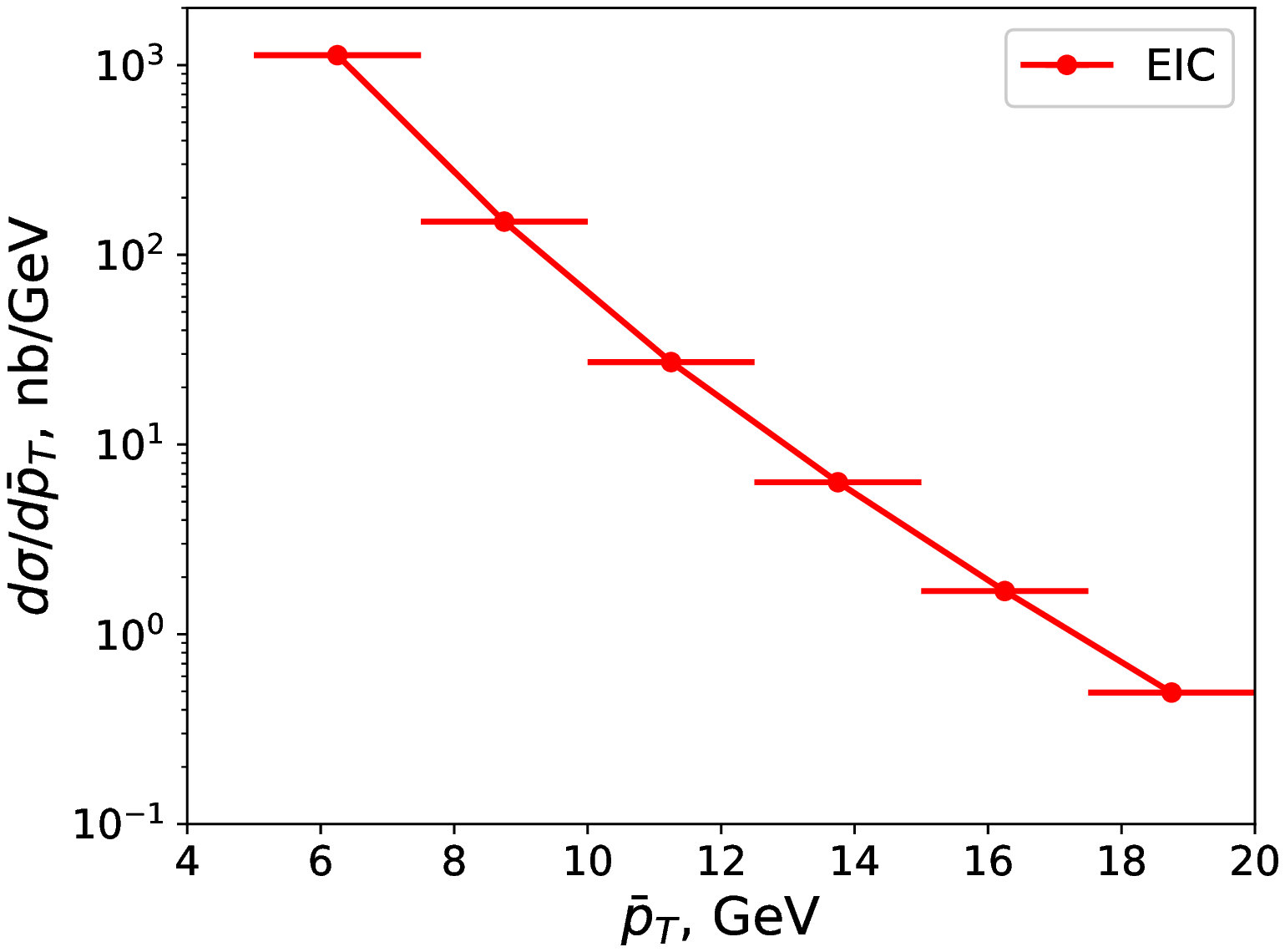}
\hspace{-0.44cm}
\includegraphics[width=6.5cm]{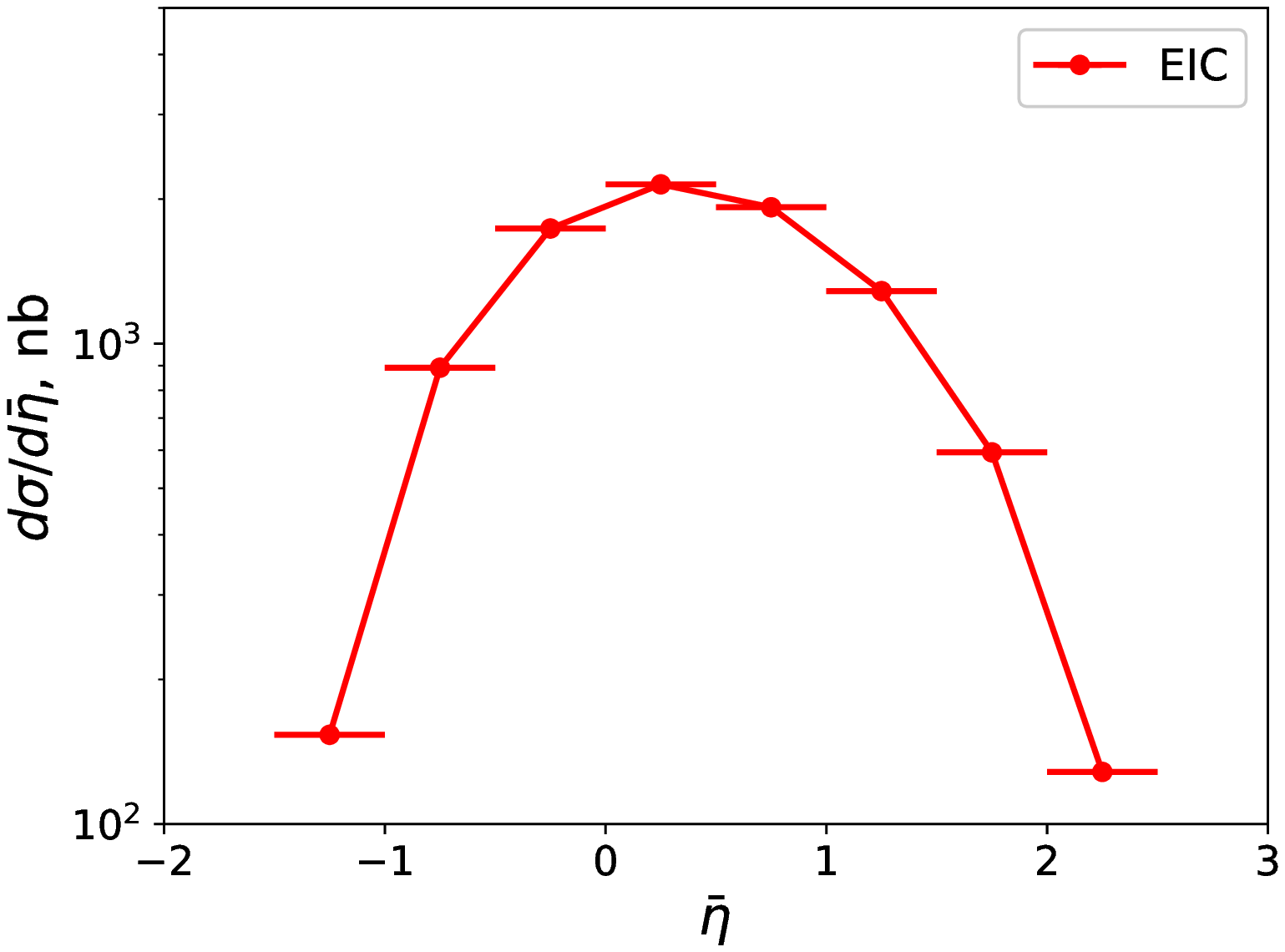}}
\centerline{
\includegraphics[width=6.5cm]{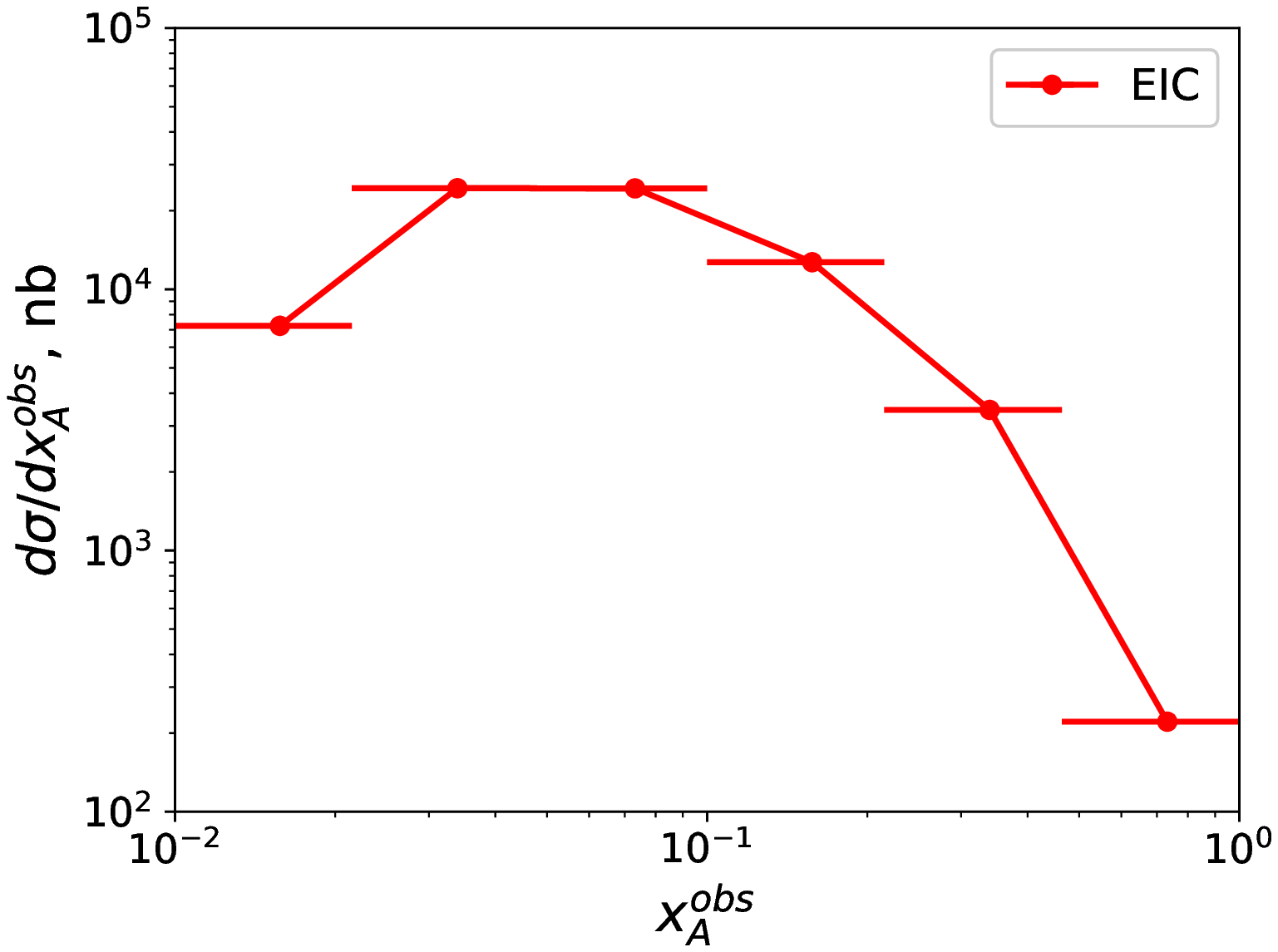}
\hspace{-0.44cm}
\includegraphics[width=6.5cm]{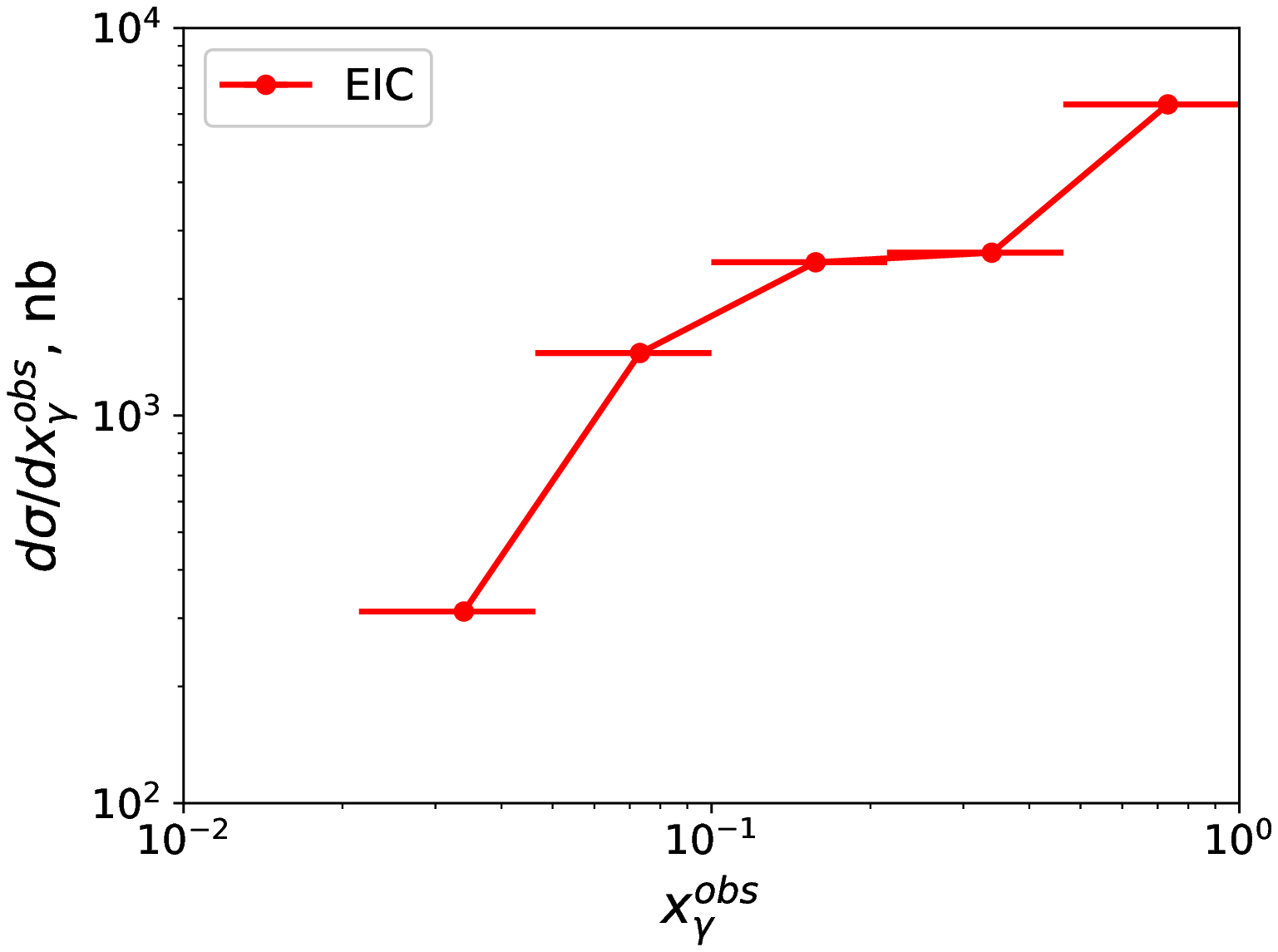}
}
\caption{NLO pQCD predictions for the $e+A \to e^{\prime}+{\rm 2\,jets}+X$ dijet photoproduction cross section in $eA$ scattering at the EIC
as a function of the average dijet transverse momentum ${\bar p}_T$, the average rapidity ${\bar \eta}$, and the momentum fractions 
$x_{A}^{\rm obs}$ and $x_{\gamma}^{\rm obs}$.}
\label{fig:res_eA_inc}
\end{figure}

The resulting distributions in the dijet average transverse momentum ${\bar p}_T=(p_{T,1}+p_{T,2})/2$, the dijet average 
rapidity ${\bar \eta}=(\eta_1+\eta_2)/2$, and the observed nucleus and photon momentum fractions, 
$x_{A}^{\rm obs}$ and $x_{\gamma}^{\rm obs}$, are shown in Fig.~\ref{fig:res_eA_inc}. One can see from the figure that at the EIC, the kinematic reach in
these variables is $5 < {\bar p}_T < 20$ GeV, $-1 < {\bar \eta} < 2$, $0.01 < x_A^{\rm obs} < 1$, and 
$0.03 < x_{\gamma}^{\rm obs} < 1$. 

Going from the EIC to LHeC and further to FCC, the collision energy increases, which subsequently dramatically expands the kinematic coverage.
In particular, it was shown in~\cite{Guzey:2020zza} that dijet photoproduction in $eA$ scattering can be probed there at
$5 < {\bar p}_T < 60$ GeV, $-2 < {\bar \eta} < 4$, $10^{-5}-10^{-4} < x_A^{\rm obs} < 1$, and $10^{-3} < x_{\gamma}^{\rm obs} < 1$.

While the nucleus momentum fraction $x_{A}^{\rm obs}$ at the EIC has a modest kinematic reach in the small-$x$ region,  
the dijet cross section is nevertheless sensitive to nuclear modifications of PDFs: the ratio of the cross sections on the nucleus and 
the proton as a function of $x_{A}^{\rm obs}$  exhibits a $10-20$\% suppression (nuclear shadowing) at small $x_{A}^{\rm obs}$  followed
by a $10-20$\% enhancement at $x_{A}^{\rm obs} \sim 0.1$ (nuclear antishadowing), which are characteristic for the gluon nuclear PDFs.
Note, however, that the magnitude of the observed effects is compatible with sizable uncertainties of the nuclear PDFs.
The similar behavior is also obtained, when we use the EPPS16 nPDFs~\cite{Eskola:2016oht} as input for our calculations.

\section{Diffractive dijet photoproduction in lepton-proton and lepton-nucleus scattering at EIC}
\label{sec:diff}

One of the major HERA physics results is the unexpected observation that diffraction makes up approximately $10-15$\% of the total electron-proton ($ep$) 
DIS cross section~\cite{Klein:2008di,Newman:2013ada}.
Similarly to the case of inclusive scattering, one can define diffractive PDFs in the framework of collinear QCD factorization~\cite{Collins:1997sr}, extract them from the HERA data on the proton diffractive 
structure functions~\cite{H1:2006zyl,ZEUS:2009uxs}, and test their universality in diffractive dijet and open charm production in DIS~\cite{H1:2007oqt,ZEUS:2007yji}. 
At the same time, it was found that NLO pQCD overestimates the measured cross section of diffractive dijet photoproduction by approximately a factor 
of 2~\cite{H1:2010xdi,H1:2015okx,ZEUS:2007uvk},
%~\cite{H1:2007jtx,H1:2010xdi,H1:2015okx,ZEUS:2007uvk}, 
which indicates breaking of the QCD factorization.
The mechanism of it remains unknown: the theory and the data can be made consistent by introducing either the global suppression factor of $R_{\rm glob}=0.5$ or the suppression factor of $R_{\rm dir}=0.34$ for the resolved-photon contribution only or 
the $x_{\gamma}$-dependent suppression factor interpolating between these two scenarios~\cite{Guzey:2016awf}.

Diffractive dijet photoproduction corresponds to the situation, when one requires that the target hadron (proton, nucleus) in Fig.~\ref{fig:photo_dijets_eA} stays intact or dissociates into a low-mass excitation. In the proton target case, 
the $e+p \to e^{\prime}+{\rm 2\,jets}+X^{\prime}+Y$ cross section of diffractive dijet photoproduction in NLO pQCD reads [compare to Eq.~(\ref{eq:cs_inc})]
\begin{eqnarray}
d\sigma(e+p &\to& e^{\prime}+ 2\,{\rm jets}+X^{\prime}+Y)=\sum_{a,b}\int dy  \int dx_{\gamma} \int dt \int dx_{\Pomeron} \int dz_{\Pomeron} f_{\gamma/e}(y) \nonumber\\
&\times& f_{a/\gamma}(x_{\gamma},\mu^2) f_{b/p}^{D(4)}(z_{\Pomeron},\mu^2,x_{\Pomeron},t) d{\hat \sigma}(ab \to {\rm jets}) \,,
\label{eq:cs_diff}
\end{eqnarray}
where $f_{b/p}^{D(4)}(z_{\Pomeron},\mu^2,x_{\Pomeron},t)$ is the so-called diffractive PDF of the proton.
It is a conditional probability to find parton $b$ with the momentum fraction $z_{\Pomeron}$ with respect to the diffractive exchange 
carrying the momentum fraction $x_{\Pomeron}$
(often called the Pomeron) provided that the final-state proton (or its low-mass excitation $Y$) receives the momentum transfer
squared $t$. To further illustrate this concept, it is convenient to assume the so-called Regge factorization for diffractive PDFs, where
they are given as product of the Pomeron flux $f_{\Pomeron}/p(x_{\Pomeron},t)$ and the PDFs of the Pomeron 
$f_{b/\Pomeron}(z_{\Pomeron},\mu^2)$,
\begin{equation}
f_{b/p}^{D(4)}(z_{\Pomeron},\mu^2,x_{\Pomeron},t)=f_{\Pomeron}/p(x_{\Pomeron},t) f_{b/\Pomeron}(z_{\Pomeron},\mu^2) 
+f_{\Reggeon}/p(x_{\Pomeron},t) f_{b/\Reggeon}(z_{\Pomeron},\mu^2) \,.
\label{eq:Regge}
\end{equation}
In Eq.~(\ref{eq:Regge}), the second term gives the sub-leading Reggeon contribution, which becomes important only for large 
$x_{\Pomeron} > 0.03$~\cite{H1:2006zyl}.

\begin{figure}[htb]
\centerline{%
\includegraphics[width=6.5cm]{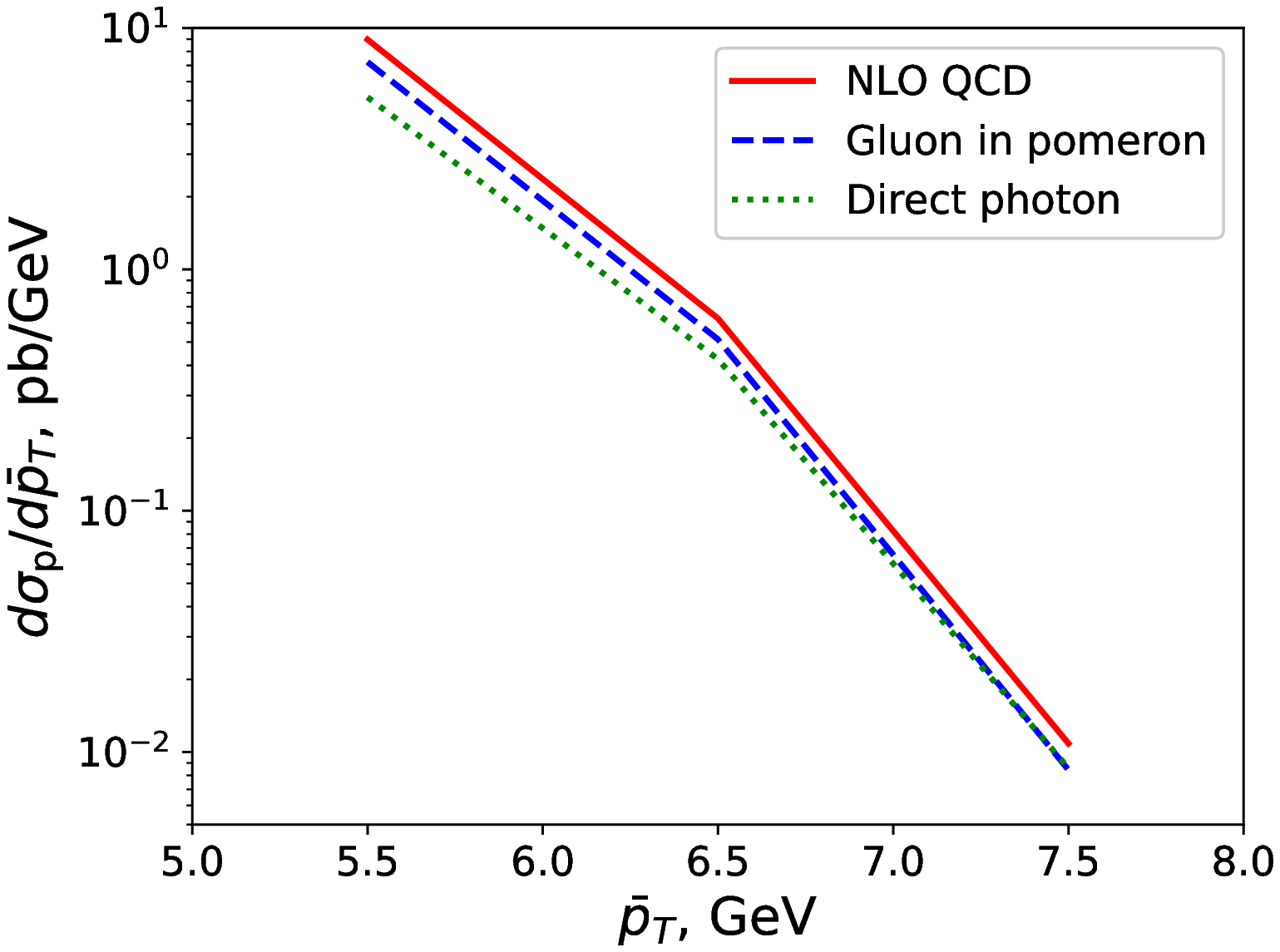}
\hspace{-0.44cm}
\includegraphics[width=6.5cm]{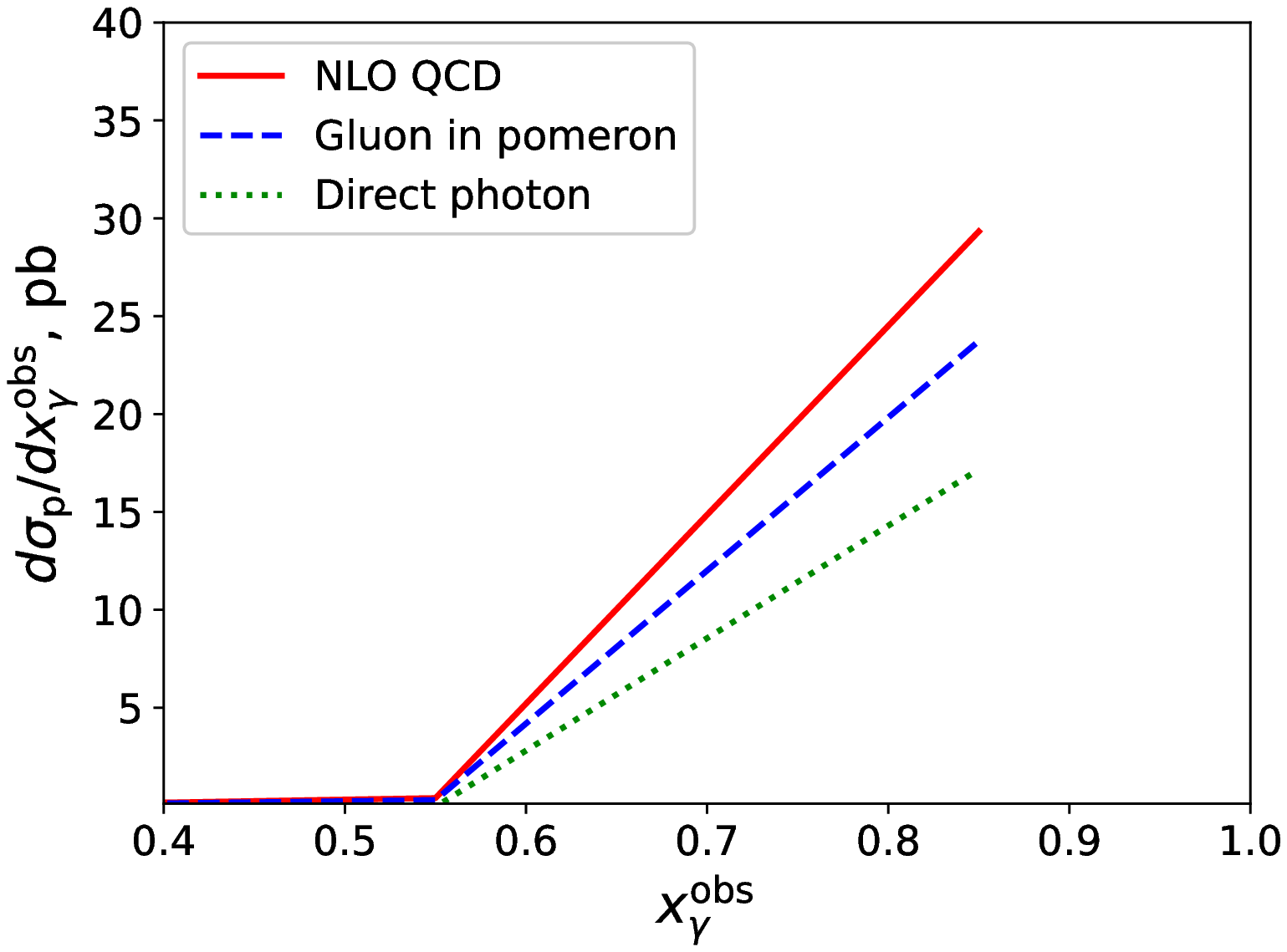}
}
\caption{NLO pQCD predictions for the $e+p \to e^{\prime}+{\rm 2\,jets}+X^{\prime}+Y$ cross section of diffractive dijet photoproduction 
in $ep$ scattering at the EIC as a function of the average dijet transverse momentum ${\bar p}_T$ and the photon momentum fraction 
$x_{\gamma}^{\rm obs}$.}
\label{fig:res_ep_diff}
\end{figure}

Using the numerical implementation of Eq.~(\ref{eq:cs_diff}) discussed above, we make predictions for diffractive dijet photoprodiction
in $ep$ scattering at the EIC~\cite{Guzey:2020gkk}. In addition to the generic cuts and the energy configuration ($E_p=100$ GeV, $E_e=21$ GeV)
) discussed in Sec.~\ref{sec:inc}, we take $|t| < 1$ GeV$^2$, $M_Y < 1.6$ GeV, and $x_{\Pomeron} \leq 0.03$
and use H1 2006 Fit B for proton diffractive PDFs~\cite{H1:2006zyl}. 

An example of our predictions is presented in Fig.~\ref{fig:res_ep_diff} showing the distributions in the dijet average 
transverse momentum ${\bar p}_T$ (left) and the photon momentum fraction $x_{\gamma}^{\rm obs}$ (right).
The red solid curves give the full result, where we use only the Pomeron contribution in Eq.~(\ref{eq:Regge}), the blue 
dashed curves show the contribution of the gluon diffractive PDF, and the green dotted curves are the direct-photon contribution.
One can see from the figure that the coverage in both ${\bar p}_T$ and $x_{\gamma}^{\rm obs}$ is rather limited. In the accessible range of 
$x_{\gamma}^{\rm obs} > 0.5$, the cross section is dominated by the contributions of direct photons and point-like quark-antiquark pairs, 
which makes it difficult to study the mechanism of factorization breaking mentioned above.
Also, the cross section probes large values of $x_{\Pomeron}$ and $z_{\Pomeron}$, which results in the dominance of the gluon 
diffractive PDF. 

\begin{figure}[htb]
\centerline{%
\includegraphics[width=6.5cm]{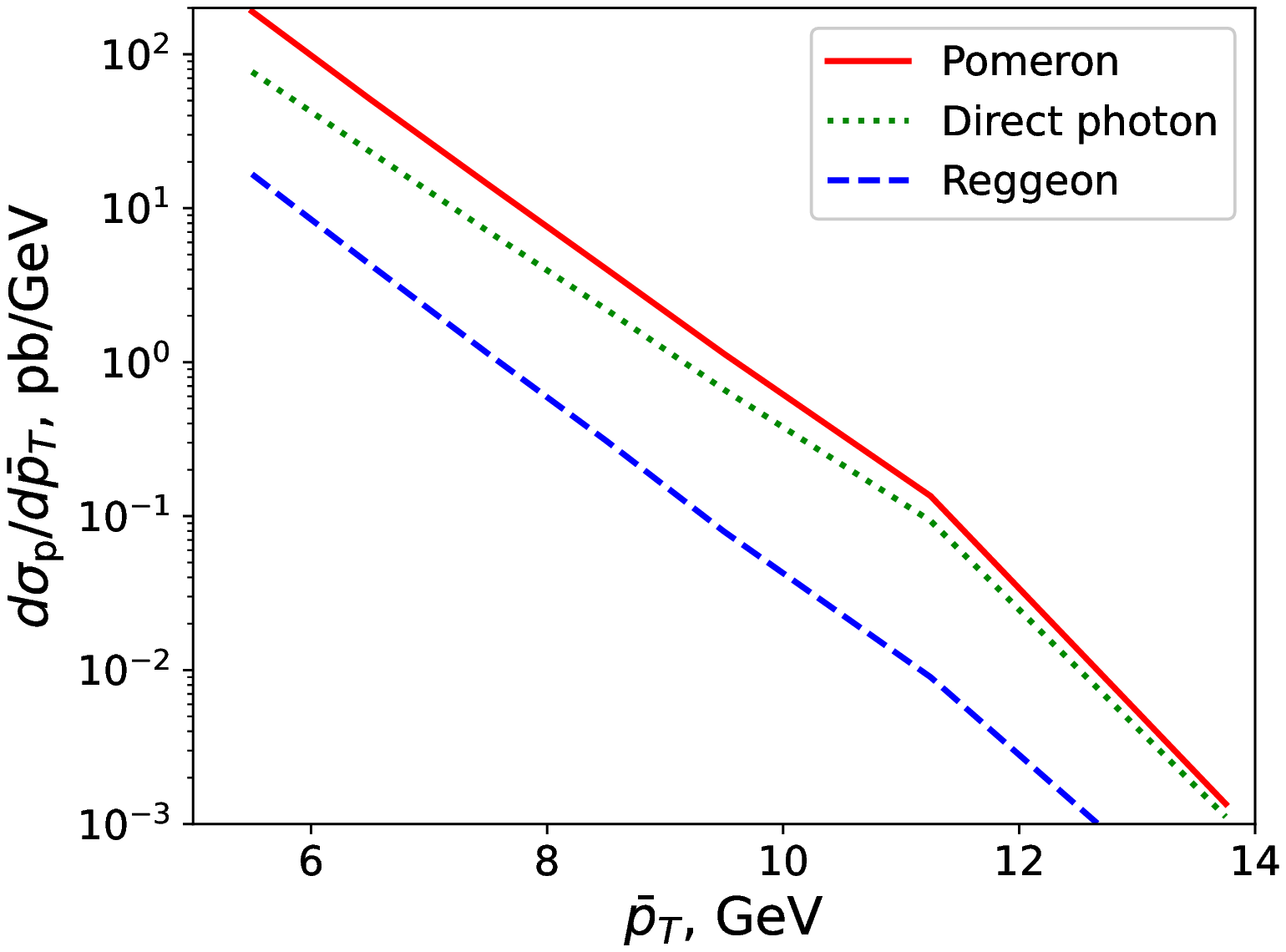}
\hspace{-0.44cm}
\includegraphics[width=6.5cm]{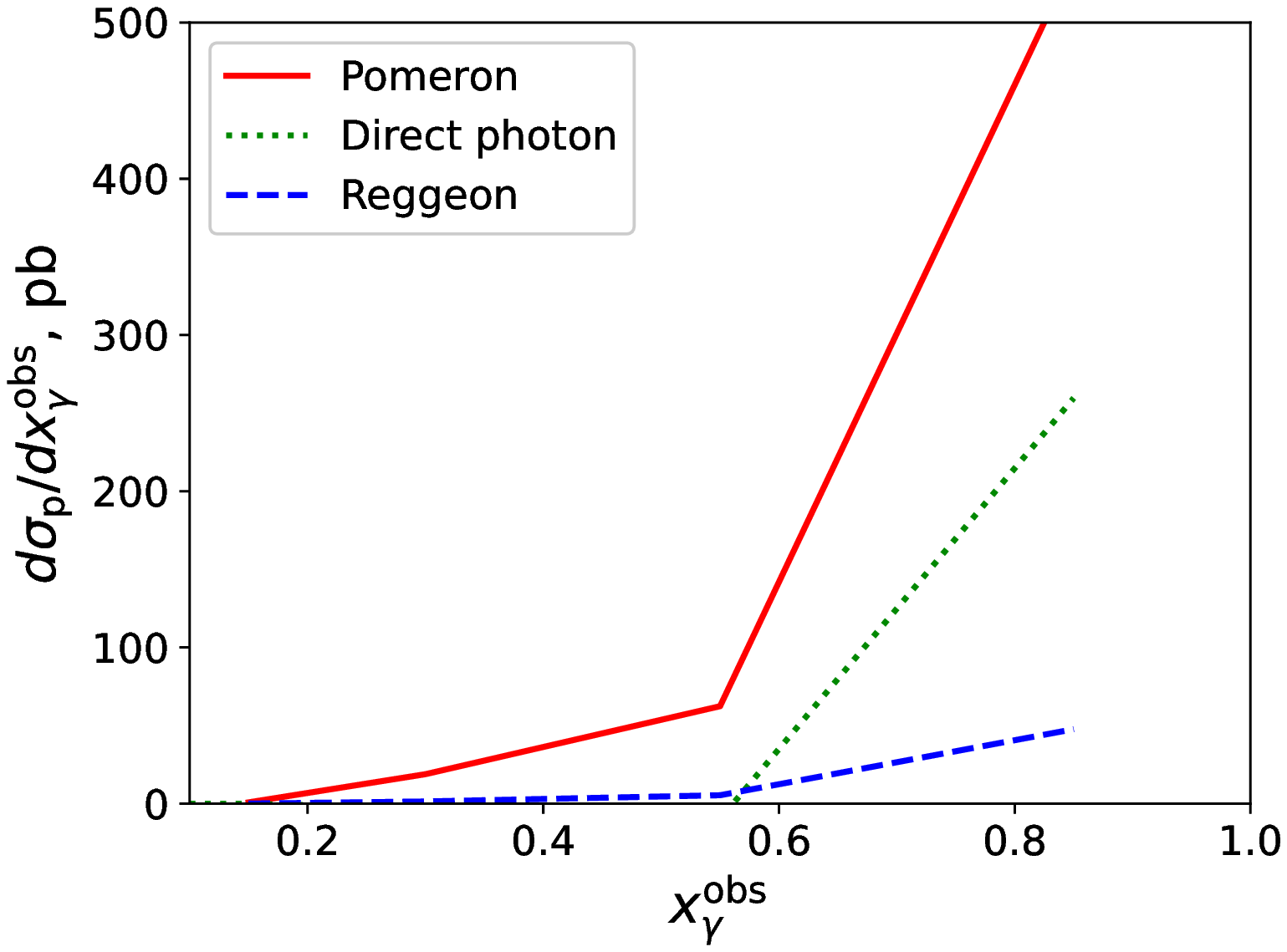}
}
\caption{Same as Fig.~\ref{fig:res_ep_diff}, but now with an extended range in $x_{\Pomeron} < 0.1$. The sub-leading Reggeon contribution is shown by the blue dashed lines.}
\label{fig:res_ep_diff_xp}
\end{figure}

To extended the kinematic coverage, we repeated our analysis using a larger range in $x_{\Pomeron}$ up to $x_{\Pomeron} < 0.1$. 
The results for the ${\bar p}_T$ and $x_{\gamma}^{\rm obs}$ distributions are presented in Fig.~\ref{fig:res_ep_diff_xp}.
The red solid and blue dashed curves correspond to the Pomeron and Reggeon contributions, see Eq.~(\ref{eq:Regge});
the green dotted curves give the direct-photon contribution. A comparison to Fig.~\ref{fig:res_ep_diff} demonstrates that 
the use of the $x_{\Pomeron} < 0.1$ range extends the coverage up to ${\bar p}_T < 14$ GeV and down to $x_{\gamma}^{\rm obs} > 0.1$.
In addition, it brings about the sub-leading Reggeon trajectory, which now contributes at the level of $10-35$\% for $x_{\Pomeron} > 0.06$.

We discussed above that NLO pQCD predictions for diffractive dijet photoproduction should be in general 
supplemented by the factor accounting for the QCD factorization breaking. Since its mechanism involves an interplay of
the direct-photon and resolved-photon contributions, the most sensitive observable is the $x_{\gamma}^{\rm obs}$ distribution.
To disentangle competing scenarios of the factorization breaking, one needs a sufficiently large range in $x_{\gamma}^{\rm obs}$, 
which is turn requires the highest proton beam energy and high precision since the cross section falls by two orders of magnitude.
Our analysis~\cite{Guzey:2020gkk} demonstrated that the assumed pattern of factorization breaking affects mostly   
the normalization of the ${\bar p}_T$ distribution (and other kinematic distributions) and only rather moderately 
the shape of the $x_{\gamma}^{\rm obs}$ distribution.

To better differentiate among different schemes of factorization breaking, one can study diffractive dijet photoproduction
in electron-nucleus ($eA$) scattering at the EIC, $e+A \to e^{\prime}+{\rm 2\,jets}+X^{\prime}+A$, where nuclei 
play the role of ``filters'' for different components of the photon wave function 
in photon-nucleus scattering. In addition, it will allow one to 
probe the novel nuclear diffractive PDFs. 

At small values of $x_{\Pomeron}$ relevant for diffraction, nuclear diffractive PDFs are expected to be suppressed
compared to their free proton counterparts due to nuclear shadowing.  In the leading twist approach~\cite{Frankfurt:2011cs}, $t$-integrated nuclear diffractive PDFs $f_{i/A}^{D(3)}(z_{\Pomeron},\mu^2,x_{\Pomeron})$
are obtained by summing the diagrams corresponding to coherent diffractive scattering on 1, 2, $\dots$, $A$ nucleons  
of the nuclear target, 
\begin{eqnarray}
f_{i/A}^{D(3)}(z_{\Pomeron},\mu^2,x_{\Pomeron}) &=&16 \pi f_{i/p}^{D(4)}(z_{\Pomeron},\mu^2,x_{\Pomeron},t=0) \nonumber\\
& \times& 
\int d^2 \vec{b} \left|\frac{1-e^{-\frac{1}{2}(1-i \eta))\sigma_{\rm soft}^i(x,\mu^2) T_A(b)}}{(1-i \eta)\sigma_{\rm soft}^i(x,\mu^2)} \right|^2
\label{eq:nuc} \,.
\end{eqnarray} 
Here $T_A(b)=\int dz \rho_A(b,z)$ is the nuclear optical density, where $\rho_A(b,z)$ is the nuclear density and 
$\vec{b}$ is the transverse position of the interacting nucleon; $\sigma_{\rm soft}^i(x,\mu^2)$ is the effective soft cross section
controlling the strength of the interaction with the target nucleons and $\eta=0.15$ is the ratio of the real to imaginary parts
of the corresponding scattering amplitude. One can see from Eq.~(\ref{eq:nuc}) that nuclear shadowing explicitly violates the Regge factorization for nuclear diffractive PDFs [compare to the proton case in Eq.~(\ref{eq:Regge})].

In practice, to estimate yields and kinematic distributions, one can use the numerical observation that the effect of nuclear shadowing in 
Eq.~(\ref{eq:nuc}) in most of the kinematics weakly depends on the parton flavor $i$, the momentum fractions $z_{\Pomeron}$ and 
$x_{\Pomeron}$, and scale $\mu$. In this case, the nuclear diffractive PDFs are given by the following simple expression,
\begin{equation}
f_{i/A}^{D(3)}(z_{\Pomeron},\mu^2,x_{\Pomeron}) = A R(x,A)f_{i/p}^{D(3)}(z_{\Pomeron},\mu^2,x_{\Pomeron}) \,,
\label{eq:nuc_approx}
\end{equation}
where $A$ is the nucleus atomic mass number and $R(x,A) \approx 0.65$ is a weak function of $x$ and $A$ calculated using
Eq.~(\ref{eq:nuc}).
Replacing proton diffractive PDFs by nuclear diffractive PDFs in Eq.~(\ref{eq:cs_diff}), one can readily make predictions for 
the  $e+A \to e^{\prime}+{\rm 2\,jets}+X^{\prime}+A$ cross section of coherent dijet photoproduction on nuclei in the EIC kinematics.

\begin{figure}[htb]
\centerline{%
\includegraphics[width=9.5cm]{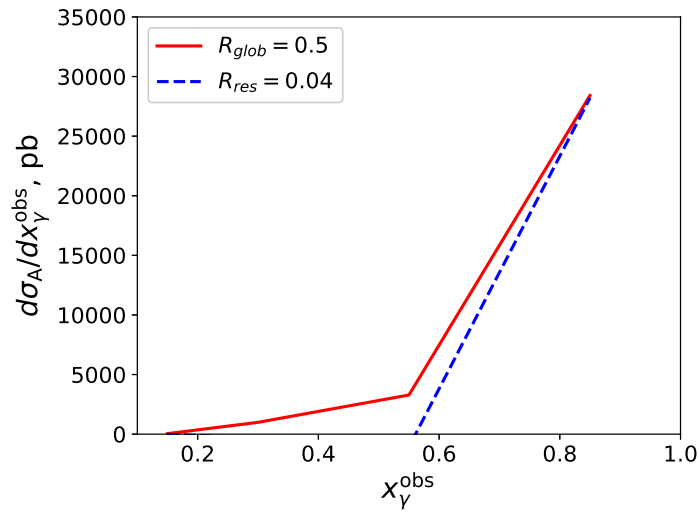}
}
\caption{NLO pQCD predictions for the $e+A \to e^{\prime}+{\rm 2\,jets}+X^{\prime}+A$ cross section of coherent diffractive dijet photoproduction on Au-197 at the EIC as a function of the photon momentum fraction 
$x_{\gamma}^{\rm obs}$. The red solid and blue dashed curves corresponds to the two assumed schemes of factorization breaking, see text for detail.}
\label{fig:res_eA_diff}
\end{figure}

Figure~\ref{fig:res_eA_diff} shows the $x_{\gamma}^{\rm obs}$ distribution for the gold nucleus (Au-197) and contrasts two scenarios of the QCD factorization breaking in diffraction: the red solid curve corresponds to the global suppression factor of $R_{\rm glob}=0.5$ as in the proton case
and the blue dashed curve is obtained by applying the $R_{\rm res}=0.04$ suppression factor to the resolved-photon contribution.
One can see from the figure that the two scenarios lead to sufficiently different predictions for $x_{\gamma}^{\rm obs} < 0.5$.

\section{Conclusions}

Photoproduction of dijets is a standard tool of QCD. Its theory is well-established in NLO pQCD, whose predictions compare very well 
to HERA data.
Inclusive and diffractive dijet photoproduction at the EIC are complimentary to respective DIS measurements
and can help constrain proton and nucleus usual and diffractive PDFs. 
In addition, diffractive dijet photoproduction at the EIC may shed some light on the outstanding problem of factorization breaking.
This requires a wide coverage in $x_{\gamma}^{\rm obs}$, which is provided by the highest proton beam energy and 
a large range in $x_{\Pomeron}$, and will benefit from the use of nuclear beams.

\section*{Acknowledgements}

The research of V.G. was funded by the Academy of Finland project 330448, the Center of Excellence in
Quark Matter of the Academy of Finland (projects 346325 and 346326), and the European Research Council project ERC-2018-ADG-835105 YoctoLHC.
The work of M.K. was also funded by the DFG through the Research Training
Group 2149 ``Strong and Weak Interactions - from Hadrons to Dark Matter''
and the SFB 1225 ``Isoquant'', project-id 273811115.

\end{document}